# High-efficiency and high-power single-frequency fiber laser at 1.6 μm based on cascaded energy-transfer pumping

XIANCHAO GUAN,[1,2] QILAI ZHAO,[2] WEI LIN,[1,2] TIANYI TAN,[2] CHANGSHENG YANG,[2,4,6,*] PENGFEI MA,[2,7] ZHONGMIN YANG,[1,2,3,5,6] AND SHANHUI XU[2,3,4]

[1]School of Physics and Optoelectronics, South China University of Technology, Guangzhou 510640, China
[2]State Key Laboratory of Luminescent Materials and Devices and Institute of Optical Communication Materials, South China University of Technology, Guangzhou 510640, China
[3]Guangdong Engineering Technology Research and Development Center of High-performance Fiber Laser Techniques and Equipment, Zhuhai 519031, China
[4]Hengqin Firay Sci-Tech Company Ltd., Zhuhai 519031, China
[5]Guangdong Engineering Technology Research and Development Center of Special Optical Fiber Materials and Devices, Guangzhou 510640, China
[6]Guangdong Provincial Key Laboratory of Fiber Laser Materials and Applied Techniques, South China University of Technology, Guangzhou 510640, China
[7]pengfeima_scut@163.com
*Corresponding author: mscsyang@scut.edu.cn

**In this paper, a technique combing cascaded energy-transfer pumping (CEP) method and master-oscillator power-amplifier (MOPA) configuration is proposed for power scaling of 1.6-μm-band single-frequency fiber lasers (SFFLs), where the $Er^{3+}$ ion has a limited gain. The CEP technique is fulfilled by coupling a primary signal light at 1.6 μm and a C-band auxiliary laser. The numerical model of the fiber amplifier with the CEP technique reveals that the energy transfer process involves the pump competition and the in-band particle transition between the signal and auxiliary lights. Moreover, for the signal emission, the population density in the upper level is enhanced and the effective population inversion is achieved due to the CEP. A single-frequency MOPA laser at 1603 nm with an output power of 52.6 W and an improved slope efficiency of 30.4% is obtained experimentally through the CEP technique. Besides, a laser linewidth of 5.2 kHz and a polarization-extinction ratio of ~18 dB are obtained at the maximum output power. The proposed technique provides an optional method of increasing the slope efficiency and power scaling for fiber lasers operating at L-band.**

## 1. INTRODUCTION

Single-frequency fiber lasers (SFFLs) operating at 1.6 μm are very potential for many applications, such as atmospheric remote sensing, optical frequency standards, free-space communication, high-resolution molecular spectroscopy and pump source for $Tm^{3+}$-doped or $Tm^{3+}$/$Ho^{3+}$ co-doped gain media [1-5]. The master laser in the injection-locked system for coherent lidar is also one of the most promising applications of the 1.6 μm SFFL with high power and narrow linewidth [6]. Single-frequency (SF) lasers operating at 1.6 μm have been achieved to generated and amplified based on $Er^{3+}$-doped YAG crystals [7-11]. Nonetheless, lasers with all-fiber structure possess unique advantages of their compactness, excellent beam quality, environmental reliability, and start to challenge the bulk solid-state lasers [12]. Hence, from the perspective of application requirements, it is desired to obtain an SFFLs operating at 1.6 μm with superior performances, such as high output power, narrow linewidth, and all-fiber structure.

A linearly-polarized SFFL at 1603 nm with an output power of 20 mW and a laser linewidth of 1.9 kHz has been successfully achieved on the foundation of the heavily $Er^{3+}$/$Yb^{3+}$ co-doped phosphate glass fiber in our group [13]. Unfortunately, for the 1.6 μm SFFLs all-fiber master-oscillator power-amplifier (MOPA) configuration, the output powers were usually limited within 20 W and slope efficiencies were around 20% in several experiments because of the low emission cross-section of $Er^{3+}$ ion near 1.6 μm [6, 14-16]. Besides, the power scaling of 1.6-μm lasers is facile to be saturated, and the signal characteristics severely degrade due to a large amount of the amplified spontaneous emission (ASE) [13, 17]. Usually, more pump power and longer active fiber can offer more gain for signal light in MOPA system. However, for SFFLs working at 1.6 μm, increasing the active fiber length makes the Brillouin gain improves easier than the signal gain and strongly limits the signal power scaling [18, 19]. In addition, more pump power would bring about the residual energy being stored in the fiber as heat. Therefore directly increasing the pump power or the active fiber length cannot amplify the laser power at 1.6 μm effectively. According to the energy level diagram of $Er^{3+}$-doped system, C-band lasers are advantageous in-band pumping sources for 1.6 μm lasers operation. As a well-developed pumping method, in-band pumping can control excess gain in active fiber by limiting the excitation level [20].

In this paper, a cascaded energy-transfer pumping (CEP) technique that employs a C-band laser (auxiliary light) as an energy transmission link is designed for improving the output power and slope efficiency of 1.6 μm SF MOPA lasers. Theoretically, a simulated model describes the transition of particles in the energy level system with the signal light at 1603 nm and the auxiliary light at 1550 nm injected simultaneously. The conjecture of gain competition and the in-band energy transmission between the signal and auxiliary lasers has been confirmed. Experimentally, by optimizing the active fiber length and the injected power of auxiliary light, the signal light in the MOPA system evolves according to the CEP technique. Finally, a 5.2-kHz-linewidth linearly-polarized SFFL operating at 1603 nm with an output power of 52.6 W and a slope efficiency of 30.4% is demonstrated. The experimental results are in good agreement with the simulated ones.

## 2. THEORETICAL ANALYSIS

In principle, the stronger emission cross-section of the $Er^{3+}$-doped fiber is mainly concentrated in the wavelength range of 1530-1560 nm, which is much higher than that in the L-band (including 1603 nm) [16]. According to the energy level diagram of $Er^{3+}$, L and C-band emission share the approximate sublevels ($^4I_{13/2a}$ and $^4I_{13/2b}$) splitting from the energy level $^4I_{13/2}$ [14-16]. Since the upper sublevel $^4I_{13/2b}$ has more particles than $^4I_{13/2a}$, $^4I_{13/2b}$ can be applied as another source of particles for lasers at 1603 nm. The in-band rapid relaxation between these two sublevels conduces to enhance the population density and achieve the population inversion at sublevel $^4I_{13/2a}$ for emission at 1603 nm. Hence, a C-band laser is chosen as the auxiliary laser and launched into the active fiber with the signal and pump lasers in the following simulation.

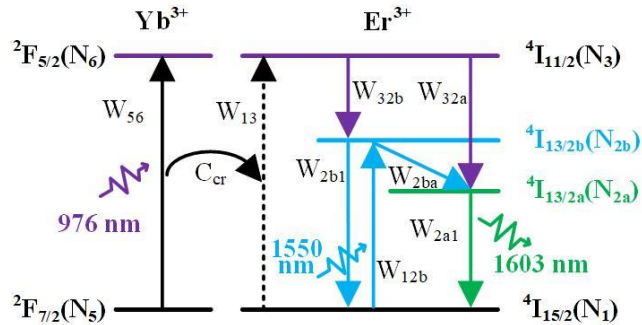

**Fig. 1.** Energy-level scheme of the $Er^{3+}/Yb^{3+}$ co-doped system (a) without and (b) with the auxiliary wave.

According to the above analysis, the physical processes therein can be described from the perspective of ion transitions. The simplified model of the $Er^{3+}/Yb^{3+}$ co-doped system is composed of a two-level structure of $Yb^{3+}$ and a three-level structure of $Er^{3+}$ [21, 22], as shown in Fig. 1. The process of power amplification in EYDF under the excitation of a 976nm LD has been studied in Ref. [22, 23]. When the 1550 nm auxiliary light is coupled into the active fiber with the signal light, it is necessary to take the split energy levels $^4I_{13/2}$ into account, namely sublevels $^4I_{13/2a}$ for signal laser and $^4I_{13/2b}$ for auxiliary laser. The 1550 nm auxiliary light has a great advantage in the gain competition with the 1603 nm laser, and more particles start to accumulate at the sublevel $^4I_{13/2b}$. After that, the particles in sublevel $^4I_{13/2b}$ transition to another sublevel $^4I_{13/2a}$ through the in-band non-radiative transition. Eventually, the power of the signal laser is amplified through stimulated emission.

According to Fig. 1, the key to the CEP technique is that the energy of the amplified 1550 nm laser is transmitted to the sub-level $^4I_{13/2b}$ by in-band fast relaxation, thereby enhancing the stimulated emission of the 1.6 μm laser. Therefore, the injected power of the auxiliary laser and the length of the active fiber mainly determine the above two processes. Based on the described model, a numerical simulation of an $Er^{3+}/Yb^{3+}$ co-doped fiber amplifier with a CEP technique is conducted [24-26]. During the simulation process, the input signal power of 2.5 W and pump power of 170 W is assumed, which are designed according to the actual conditions in our experiment. The related parameters and the physical meaning used in the simulation are listed in Table 1.

**Table 1. Related parameters used in the theoretical simulation**

| Symbol/unit | Physical meaning | Value |
|---|---|---|
| $N_{Er}$/m$^{-3}$ | $Er^{3+}$ concentrations | $3\times10^{25}$ |
| $N_{Yb}$/m$^{-3}$ | $Yb^{3+}$ concentrations | $6\times10^{25}$ |
| $A_{eff}$/m$^2$ | Effective area of doped core | $4.91\times10^{-10}$ |
| $\sigma_{12a}$/m$^2$ | Absorption cross section at 1603 nm | $2.18\times10^{-25}$ |
| $\sigma_{2a1}$/m$^2$ | Emission cross section at 1603 nm | $4.66\times10^{-25}$ |
| $\sigma_{12b}$/m$^2$ | Absorption cross section at 1550 nm | $5.13\times10^{-25}$ |
| $\sigma_{2b1}$/m$^2$ | Emission cross section at 1550 nm | $6.59\times10^{-25}$ |
| $\sigma_{13}$/m$^2$ | Absorption cross section at 976 nm | $1.68\times10^{-25}$ |
| $\sigma_{56}$/m$^2$ | Absorption cross section at 976 nm | $1.51\times10^{-24}$ |
| $\tau_{Er}$/s | Fluorescence lifetimes of $Er^{3+}$ ions | $7\times10^{-3}$ |
| $\tau_{Yb}$/s | Fluorescence lifetimes of $Yb^{3+}$ ions | $1\times10^{-3}$ |
| $C_{tr}$/m$^3\cdot$s$^{-1}$ | Cumulative up-conversion energy transfer coefficient | $0.85\times10^{-22}$ |
| $C_{cr}$/m$^3\cdot$s$^{-1}$ | Energy transfer coefficient from $^2F_{5/2}$ level to $^4I_{15/2}$ level | $2.1\times10^{-22}$ |

Figures 2(a) and (b) show the output auxiliary and signal laser powers against the active fiber length of 0-10 m with different input auxiliary powers of 0, 200, 400, 500, and 600 mW, respectively. The inset graph of Fig. 2(a) depicts the residual pump power against the active fiber length.

On the one hand, the amplification of 1550 nm auxiliary light is the first step in achieving the CEP process. The choice of the auxiliary power determines the amplification result of the signal light. It can be found from the inset graph of Fig. 2(a) that a 5-m-long active fiber is enough to absorb the pump power completely. According to Figs. 2, there is little ASE power at 1.5 μm-band generated and the maximum signal power is 18.11 W when no auxiliary light is injected into the active fiber. Once the auxiliary power is coupled into the active fiber from 200 to 500 mW, the signal power has been significantly improved. Further increase of auxiliary power to 600 mW plays an unobvious role in the improvement of the signal power. Depending on the simulated results in Figs. 2, the input auxiliary power of 500 mW is the optimal condition for CEP technique and can be used as a reference for the subsequent experiment.

On the other hand, the CEP process places a requirement on the length of the gain fiber for the integrity of the CEP process. According to Figs. 2, as the active fiber length increases, the auxiliary powers improve rapidly and reach the maximum with the ~3.1-m-long active fiber. After that, the auxiliary powers are quickly absorbed and there is almost no remained with the 5.8-m-long active fiber. The signal power after the 5.6-m-long active fiber is saturated because the energy provided by the residual auxiliary light is not sufficient to offset its loss and absorption. However, the 5.8-m-long active fiber is capable of absorbing residual auxiliary light. Hence, the 5.6-m-long active fiber ensures that the CEP process described in Fig. 1 is complete and the maximum signal power is obtained. However, a 5.8-m-long active fiber is a more preferred choice to absorb the residual pump and auxiliary lights.

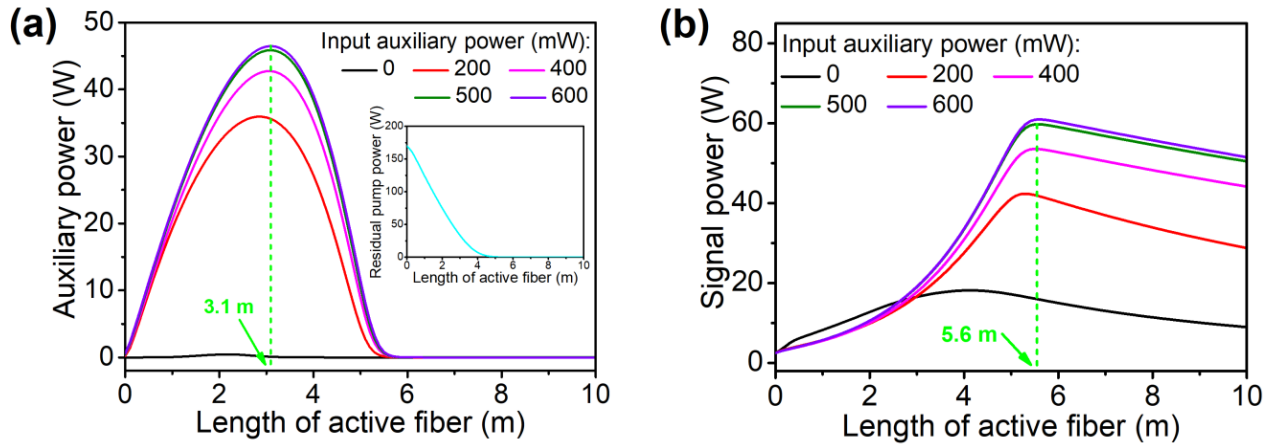

**Fig. 2.** Simulated results on the output powers of (a) auxiliary (1550 nm), inset: pump (976 nm), and (b) signal (1603 nm) waves versus the active fiber length with input auxiliary powers of 0, 200, 400, 500, and 600 mW, respectively.

Figure 3 reveals the power evolutions of the signal laser, the auxiliary laser, and the pump laser along a 5.8-m-long active fiber with the input auxiliary power of 500 mW, and elaborates the interaction among the three waves. The power evolutions of the three lasers could be divided into three stages. It can be found that the auxiliary power enhances more rapidly than that of the signal power in the front of the active fiber and achieves the maximum power first in stage $\alpha$. Because in this stage, more pump energy is delivered to the auxiliary light (1550 nm), which has a higher emission cross-section. During stage $\beta$, the power of 1603 nm laser ramps up in an increasing slope, because the signal laser in this stage is pumped by the pump and amplified auxiliary lasers simultaneously until the pump laser almost exhausts. Ultimately, in stage $\gamma$, the signal power ascends by extracting the stored energy of 1550 nm laser, and gradually saturates by acquiring the balance between the gain and dissipation.

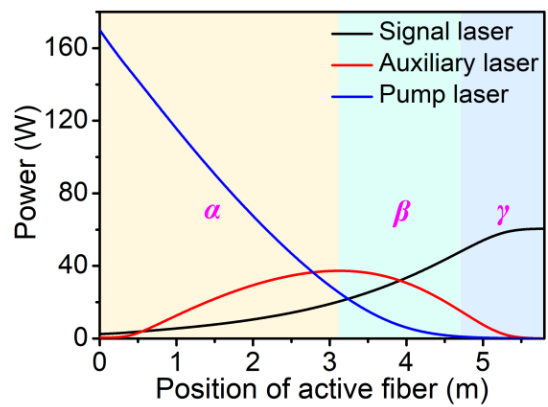

**Fig. 3.** Power evolutions of the signal, auxiliary and pump lasers in a 5.8-m-long active fiber with the input auxiliary power of 500 mW.

## 3. EXPERIMENTAL SETUP

As a demonstration of the CEP technique, a narrow-linewidth high-power polarization-maintaining (PM) single-frequency all-fiber MOPA at 1603 nm with an auxiliary laser is established. The MOPA consists of a PM distributed Bragg reflector (DBR) seed oscillator and a three-cascaded fiber amplifier, as shown in Fig. 4. The 1.6-μm DBR seed laser emits an output power of 14 mW, a laser linewidth of 5.3 kHz and a polarization-extinction ratio (PER) of 20 dB, which is similar to our previous work [13]. The signal power from the seed laser is then amplified to 190 mW and 2.5 W by two pre-amplifiers (1st pre-amplifier and 2nd pre-amplifier) with 10/128-μm core/cladding-diameter active fiber, respectively. A laser linewidth remains unchanged at 5.2 kHz and a PER is 20 dB after these two pre-amplifiers.

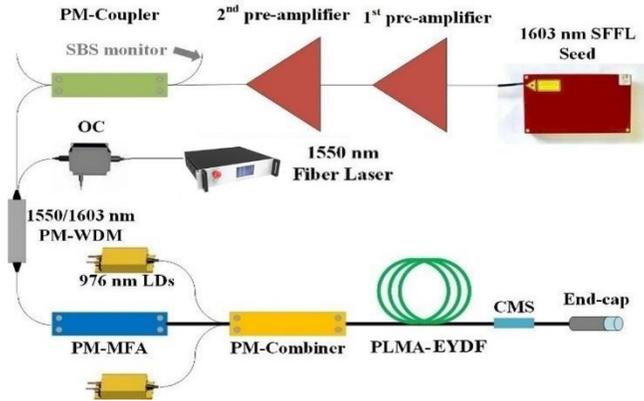

**Fig. 4.** Experimental setup of high-power narrow-linewidth PM single-frequency MOPA system at 1.6 μm.

After the second pre-amplifier, a 1550/1603 nm PM-WDM is employed to combine the homemade 1550 nm auxiliary light and the signal light. An optical circulator (OC) is used between the power-amplifier and 1550 nm fiber laser to prevent the auxiliary laser being destroyed by backward propagating 1.5-μm-band ASE. A PM large mode area $Er^{3+}/Yb^{3+}$ co-doped double cladding fiber (PLMA-EYDF) with a core/cladding diameter of 25/300 μm is employed in the power-amplifier. The active fiber is wound into a circle with a diameter of 11-15 cm to filter out higher order modes. A PM mode field adaptor (PM-MFA) is used to reduce the coupling loss between the different passive fibers. The power-amplifier is co-pumped by two 976 nm multimode LDs (total maximum power of ~170 W). A cladding-mode stripper (CMS) is utilized to remove the residual pump light. An endcap is used as the output end of the laser system to avoid any end-face reflection.

## 4. EXPERIMENTAL RESULTS AND DISCUSSION

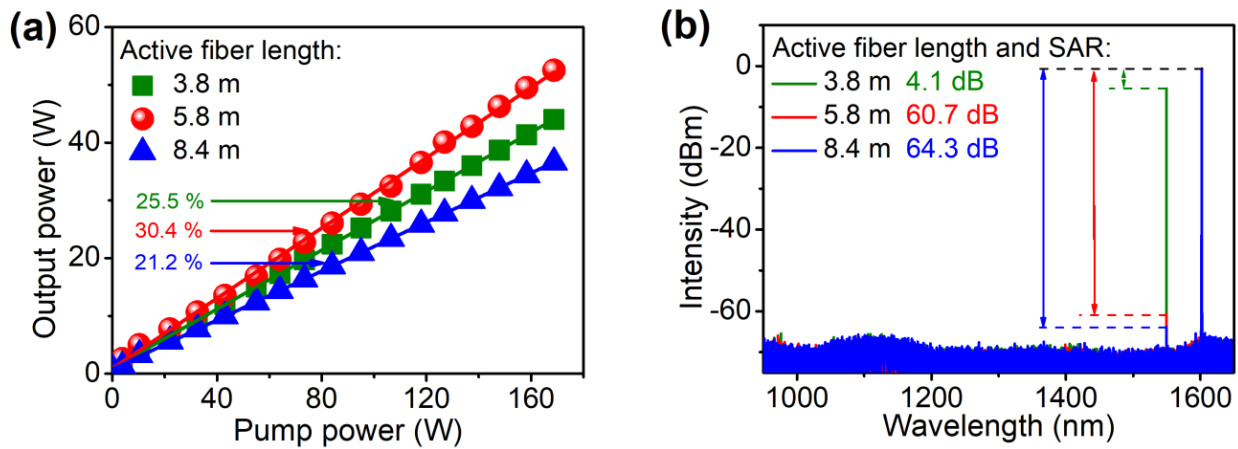

**Fig. 5.** (a) Output signal powers versus the pump power and (b) output spectra with 3.8, 5.8, and 8.4-m-long active fiber and injected auxiliary power of 500 mW.

Firstly, different lengths of active fiber are utilized to explore the impact on the laser output characteristics. The input signal power and auxiliary power are set as 2.5 W and 500 mW, respectively. Based on the above simulation results, a 5.8-m-long active fiber is selected for the power-amplifier, and the other fiber lengths of 3.8 and 8.4 m are also employed severally for comparison. The measured output powers versus the pump power are exhibited in Fig. 5(a). It can be noted that all the output powers enhance linearly with the increasing pump power. The obtained maximum signal power and the slope efficiency is 52.6 W and 30.4%, respectively, with a 5.8-m-long active fiber. By contrast, the output powers reduce to 44 and 36 W with 3.8 and 8.4-m-long ones, respectively. Moreover, both of the corresponding slope efficiencies are also lower than that with 5.8-m-long one.

Besides, the corresponding output spectra are also measured by an optical spectrum analyzer with the resolution of 0.02 nm and scanning wavelength range of 950-1650 nm at the maximum signal powers, as shown in Fig. 5(b). Here, the signal to the auxiliary-laser ratio (SAR) is defined to calibrate the component ratio of the signal light. While a 5.8-m-long active fiber is used, the SAR of 60.7 dB is obtained, and that is 64.3 dB with an 8.4-m-long one because longer active fiber absorbs more auxiliary wave. However, when a 3.8-m-long active fiber is utilized, the corresponding SAR is only 4.1 dB. According to the experimental results, short active fiber length could not provide enough gain for the signal laser and fully absorb the auxiliary wave, and long one would absorb not only the auxiliary power but also the signal power in the extra fiber, in which the trend is consistent with theoretical results in Fig. 2. Thus, according to both of the output power and spectra results, the active fiber length of 5.8 m is an optimal one.

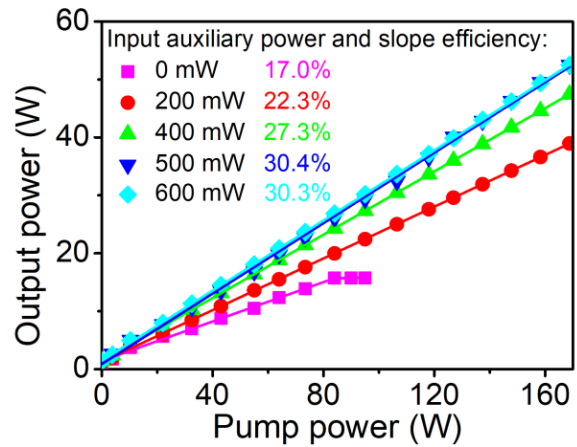

**Fig. 6.** Output powers versus the pump power with the input auxiliary power of 0, 200, 400, 500, and 600 mW, respectively.

For verifying the effect of the input auxiliary powers on the characteristics of the output wave, the output powers are measured with different input auxiliary powers from 0 to 600 mW and a 5.8-m-long active fiber, as shown in Fig. 6. Without the auxiliary laser injected (0 mW), the maximum output signal power is ~16 W and the corresponding slope efficiency of ~17%. Besides, the output signal power remains substantially unchanged even with more pump power. However, once the auxiliary laser is coupled into the active fiber with the signal wave, the slope efficiency improves to over 20%. Meanwhile, the output powers enhance linearly to more than 40 W, and no power saturation is observed. Ultimately, the maximum slope efficiency of ~30.4% and the output power of ~52.6 W, which is limited by the pump power, are obtained, when the input auxiliary power reaches 500 mW. It can be found that further improvement in the input auxiliary power has no evident contribution to the output power.

The longer fiber in MOPA always brings about a lower SBS threshold, especially for narrow-linewidth lasers. According to the SBS threshold expression [27], the SBS threshold power of the 7.3-m-long fiber amplifier (5.8-m-long active fiber and 1.5-m-long passive fiber) with a fiber core diameter of 25 μm is estimated to be 36.0 W. However, the output power of 52.6 W is actually higher than the theoretical SBS threshold. Due to the gain competition, the output power scaling of the signal laser is limited in the front portion of the active fiber, as shown in Fig. 2. Through seeding both SF signal and broadband auxiliary light, the effective length of the amplifier is shortened as the signal experiences a rapid rise at the output end of the active fiber [28]. Moreover, due to gain competition, steep thermal gradients are optically induced near the output end of the fiber, and the SBS threshold is enhanced [28]. The output signal power could be further improved through heightening more pump powers. Hence, the proposed MOPA system with a CEP technique is an effective method to enhance the SBS threshold and improve the output power scaling of a narrow-linewidth SFFL.

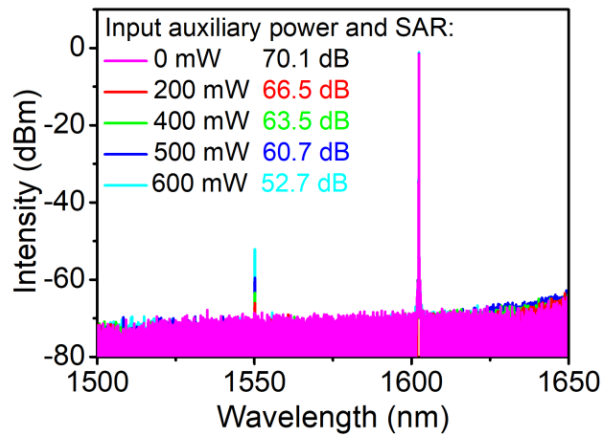

**Fig. 7.** Output spectra with the input auxiliary power of 0, 200, 400, 500, and 600 mW, respectively.

The SARs of output laser at the maximum output powers can be obtained from the measured spectra, as shown in Fig. 7. It can be seen that there is no obvious ASE left. The SAR is reduced from 70.1 to 52.7 dB with the increase of the input auxiliary power from 0 to 600 mW. Namely, higher input auxiliary power causes the greater ratio of the residual laser at 1550 nm. When the input auxiliary power is more than over 500 mW, even though the maximum output power does not continue to grow, the corresponding SAR will continue to deteriorate.

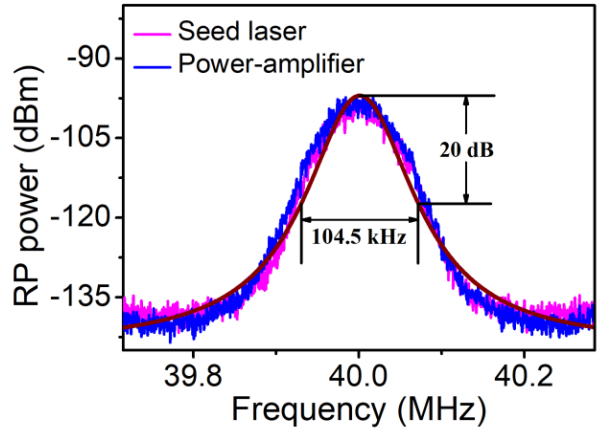

**Fig. 8**. Measured spectral linewidths at the maximum output powers.

The spectral linewidths of output laser are measured at the maximum output powers to explore whether injected auxiliary laser would have an impact on the output characteristics, as shown in Fig. 8. For the consistency and accuracy of the experimental measurement results, when linewidths are measured, the powers injected into the photodetectors are attenuated to be the same level. The linewidth of the output laser is tested based on the self-heterodyne method. The measuring device is mainly composed of a Mach-Zehnder interferometer, a 48.8 km fiber delay line, and a 40 MHz fiber-coupled acoustic-optic modulator [28-30]. According to Fig. 8, the MOPA laser has the almost same spectral linewidth as that of the seed laser. The measured self-heterodyne signal is fitted to a Lorentzian profile to estimate the spectral linewidth with 104.5 kHz at -20 dB from the peak, indicating that the measured full width of half-maximum is 5.2 kHz.

Furthermore, the output power at the maximum is measured for more than 1 hour in Fig. 9, and the result shows that the power stability is less than ±1.4%. The inset (1) in Fig. 9 shows the measured far-field beam profile at the maximal operation power. Due to the coiled active fiber, this laser can easily work in fundamental mode. Additionally, the SF characteristic of the MOPA is verified by a scanning Fabry-Perot interferometer with a resolution of 7.5 MHz and a free spectral range of 1.5 GHz. The measurement result is shown in the inset (2) of Fig. 9, which demonstrates only one longitudinal mode oscillated stably within a scanning period.

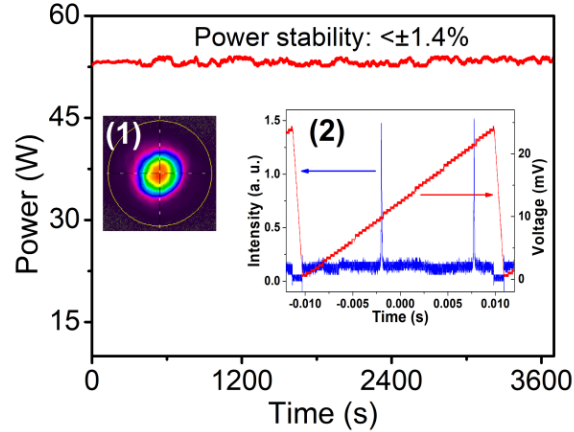

**Fig. 9.** Output power stability at the full power for >1 hour. Inset: (1) transverse shape of the output beam; (2) longitudinal-mode characteristic of the MOPA.

## 5. CONCLUSION

In conclusion, we have theoretically and experimentally demonstrated that the CEP technique could significantly improve the output power, slope efficiency and SBS threshold of the SF MOPA operating at 1603 nm. Theoretically, by injecting C-band auxiliary laser into the primary signal laser at 1.6 μm, gain competition and in-band energy-transfer are strengthened between these two bands. This MOPA system based on the CEP technique overcomes the low gain of $Er^{3+}$-doped fiber at 1.6 μm. Experimentally, a linearly-polarized SFFL operating at 1603 nm with a slope efficiency of 30.4% and maximum output power of 52.6 W is realized with the active fiber length of 5.8 m and injected auxiliary power of 500 mW. Meanwhile, a linewidth of 5.2 kHz and a SAR of 60.7 dB are obtained at the maximal output power. Further power scaling can be achieved with more pump power. The results show that this MOPA system with the CEP technique can be employed as a potentially crucial approach to the amplifier at the L-band laser, especially for narrow-linewidth SFFL.

**Funding.** National Key Research and Development Program of China (2017YFF0104602), Major Program of the National Natural Science Foundation of China (61790582), NSFC (61635004, 11674103, 61535014 and 51772101), Guangdong Key Research and Development Program (2018B090904001 and 2018B090904003), Local Innovative and Research Teams Project of Guangdong Pearl River Talents Program (2017BT01X137), Guangdong Natural Science Foundation (2016A030310410 and 2017A030310007), the Science and Technology Project of Guangdong (2016B090925004 and 2017B090911005), and the Science and Technology Project of Guangzhou (201804020028).